\documentclass[twocolumn,english,aps,prb]{revtex4-1}
\usepackage{color}
\usepackage{amsmath}
\usepackage{graphicx}
\usepackage{amssymb}
\usepackage{hyperref}
\usepackage{dsfont}

\makeatletter
\@ifundefined{textcolor}{}
{%
 \definecolor{BLACK}{gray}{0}
 \definecolor{WHITE}{gray}{1}
 \definecolor{RED}{rgb}{1,0,0}
 \definecolor{GREEN}{rgb}{0,1,0}
 \definecolor{BLUE}{rgb}{0,0,1}
 \definecolor{CYAN}{cmyk}{1,0,0,0}
 \definecolor{MAGENTA}{cmyk}{0,1,0,0}
 \definecolor{YELLOW}{cmyk}{0,0,1,0}
 }

\@ifundefined{definecolor}
 {\usepackage{color}}{}
\@ifundefined{definecolor}{\@ifundefined{definecolor}
 {\usepackage{color}}{}
}{}

\newcommand{\sgn}{\operatorname{sgn}}
\newcommand{\w}{\omega}

\newcommand{\ba}{\begin{eqnarray*}}
\newcommand{\ea}{\end{eqnarray*}}
\newcommand{\baa}{\begin{eqnarray}}
\newcommand{\eaa}{\end{eqnarray}}
\newcommand{\bea}{\begin{eqnarray}}
\newcommand{\eea}{\end{eqnarray}}
\newcommand{\be}{\begin{equation}}
\newcommand{\ee}{\end{equation}}

\newcommand{\sm}{SmB$_6$}

\newcommand{\tc}{v_c}
\newcommand{\tf}{v_f}
\newcommand{\tfb}{\bar v_f}
\newcommand{\tfbi}{\bar v_{f,i}}
\newcommand{\rhotip}{\rho_{\rm tip}}

\newcommand{\bk}{\mathbf{k}}

\newcommand{\br}{\mathbf{r}}

\DeclareMathOperator{\I}{Im}

\makeatother

\usepackage{babel}

\begin{document}

\title{
Cotunneling into a Kondo lattice with odd hybridization
}

\author{Pier Paolo Baruselli}
\author{Matthias Vojta}
\affiliation{Institut f\"ur Theoretische Physik, Technische Universit\"at Dresden, 01062 Dresden, Germany}


\begin{abstract}
Cotunneling into Kondo systems, where an electron enters a $f$-electron material via a cotunneling process through the local-moment orbital, has been proposed to explain the characteristic lineshapes observed in scanning-tunneling-spectroscopy (STS) experiments.
Here we extend the theory of electron cotunneling to Kondo-lattice systems where the bulk hybridization between conduction and $f$ electrons is odd under inversion, being particularly relevant to Kondo insulators. Compared to the case of even hybridization, we show that the interference between normal tunneling and cotunneling processes is fundamentally altered: it is entirely absent for layered, i.e., quasi-two-dimensional materials, while its energy dependence is strongly modified for three-dimensional materials.
We discuss implications for STS experiments.
\end{abstract}

\date{\today}

\pacs{}

\maketitle

\section{Introduction}

Scanning tunneling spectroscopy has developed into a powerful tool to investigate the surface electronic structure of both bulk and thin-film materials. It provides a spatially resolved map of the local single-particle density of states. Moreover it has been used to derive information on the momentum-resolved bandstructure via so-called quasiparticle interference (QPI), i.e., energy-dependent Friedel oscillations caused by dilute impurities.\cite{eigler,Hoffman1148,davis_science}
Recently, such QPI analysis has been employed\cite{seo_yazdani,yazdani,kapitulnik,xue,xia,yazdani11} to confirm the spin--momentum locking of surface states of topological insulators.\cite{Fu2007,tirev1,tirev2}

In multiorbital systems, the proper interpretation of tunneling spectra requires knowledge about individual orbital tunneling matrix elements (or their ratios) which are determined by the wavefunction overlaps between electronic states in the tip and Wannier states in the material.
For Kondo systems,\cite{hewson} involving both weakly correlated conduction electrons and local-moment $f$ electrons, it has been proposed\cite{coleman09,morr10} that both contribute to STS spectra: In addition to direct tunneling into conduction states, \textit{cotunneling} processes via localized states can occur which involve a magnetic interaction between the tunneling electron and the local moment. The quantum-mechanical interference between these two tunneling paths, together with the particle--hole (p-h) asymmetry of the conduction band, leads to a characteristic Fano-like lineshape in the STS signal. \cite{li_schneider_kondo,madhavan98,uysaghy, schiller,knorr_kern_kondo,fano_lattice,fano_lattice_uru,hf_qckl}
The interference term itself is determined by the effective coupling between $f$ and conduction electrons.

Importantly, frequently used model Hamiltonians assume a spatially local Kondo interaction or hybridization between local moments and conduction electrons, such that, for Kondo-lattice systems with inversion symmetry, the effective momentum-space hybridization function is even under inversion.
However, in real materials the Kondo interaction is not local, with its spatial structure being determined by the overlap of the relevant orbitals. In fact, in many inversion-symmetric Kondo-lattice systems the hybridization is odd under inversion -- this applies in particular to topological Kondo insulators\cite{tki1,takimoto,lu_smb6_gutz,tki_cubic} (TKIs) which have been in the focus of intense activities recently. However, the interplay of odd hybridization and cotunneling have not been studied before.

It is the purpose of this paper to study the effects of odd hybridization for cotunneling and STM spectra of Kondo-lattice systems. As we will show, odd hybridization forces the on-site propagator connecting conduction and $f$ electrons to vanish, which implies that the interference between normal tunneling and cotunneling is absent in the simplest tunneling models. Complications arise from broken inversion symmetry near the surface as well as from non-local tunneling processes, and we will discuss the consequences for STS spectra.


\section{Tunneling conductance for heavy-fermion materials}
To set the stage, we review the cotunneling theory of Refs.~\onlinecite{coleman09,morr10}. The bulk system is described by operators for conduction electrons and $f$ states, $c_{i\sigma}$ and $\bar f_{i\sigma}$, respectively; explicit Hamiltonians follow below. A spatially local tunneling process between tip and surface is captured by a Hamiltonian $\mathcal{H}_T = \sum_\sigma p_\sigma^\dagger \psi_{0\sigma} + h.c.$ where $p$ is an electron operator for the tip, and $\psi_0$ involves both $c$ and $f$ electrons at site $0$ below the tip. Importantly, Coulomb interactions on $f$ orbitals are strong and suppress charge fluctuations, leading to cotunneling. In a Kondo-model description, $f$ electrons form local moments $\vec S_i$, and $\psi_0$ can be expressed as \cite{coleman09}
\begin{equation}
\psi_{0\sigma} = \tc c_{0\sigma} + (1/2) \sum_i \tfbi (\vec\tau_{\sigma\sigma'} \cdot \vec S_0)c_{i\sigma'}\,.
\end{equation}
Here, $\tc$ and $\tfb$ are the amplitudes for direct tunneling and cotunneling, respectively, $\vec\tau$ is a Pauli matrix, and $i$ represent the $c$ electron site(s) which hybridize with the $f$ orbital at $0$. In an alternative Anderson-model description, the $f$ electrons become renormalized by Coulomb interactions, which can be captured by a constant renormalization factor $b$ at the mean-field level, $f_{i\sigma} = b \bar f_{i\sigma}$, with $f_{i\sigma}$ being auxiliary fermion operators.\cite{hewson} As a result, the operator $\psi_0$ takes the form\cite{coleman09}
\begin{equation}
\psi_{0\sigma} = \tc c_{0\sigma} + \tf f_{0\sigma}\,.
\label{psicf}
\end{equation}
The same renormalized single-particle formulation can be obtained from the Kondo description, employing a SU($N$) large-$N$ limit and a mean-field decoupling of the Kondo interaction,\cite{hewson} leading to
$\tf = -\sum_i \langle f_{0\sigma}^\dagger c_{i\sigma} \rangle \tfbi$.

\begin{figure}[tb]
\includegraphics[width=0.4\textwidth]{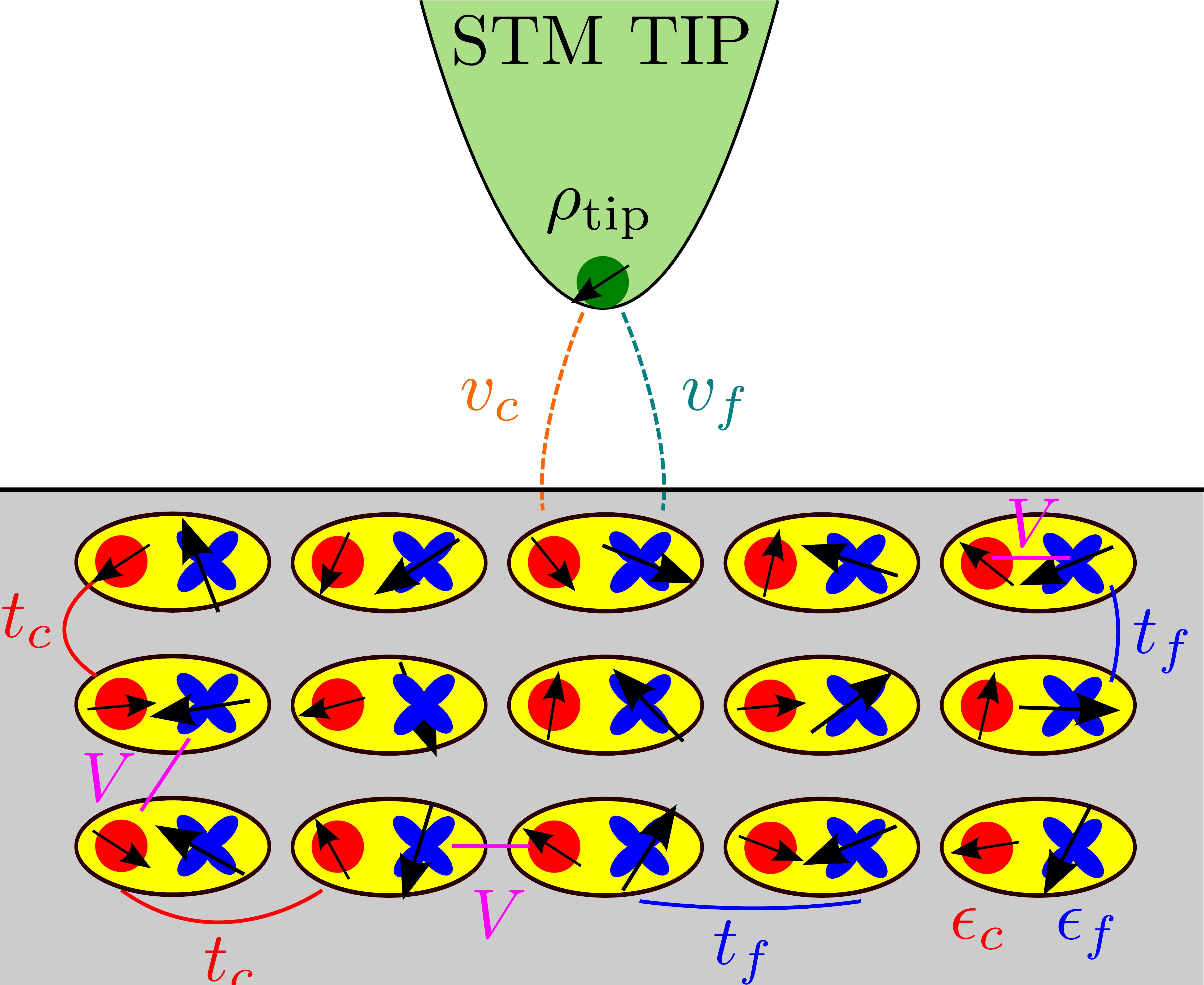}
\caption{
Schematic picture of cotunneling into a 3D heavy-fermion material.
Each site (yellow) consists of a conduction $c$ (red) and a localized $f$ (blue) orbital.
$\epsilon_c$ describes the on-site $c$ energy, $t_c$ the NN $c$ hopping,
$\epsilon_f$ the on-site $f$ energy, $t_f$ the NN $f$ hopping,
and $V$ the NN $c-f$ hybridization, which can be local (top right), or non-local (bottom left).
The STM tip is described by its DOS $\rhotip$.
Electron tunnel vertically from the tip into a $c$ orbital with amplitude $\tc$ or into a (renormalized) $f$ orbital with amplitude $\tf$.
}\label{fig_stm}
\end{figure}

Given $\mathcal{H}_T$, the tunneling conductance at voltage $\Phi$ is
\be
g(e\Phi)=-\frac{4 e^2}{\hbar} \rhotip \I [G_{\psi}(e\Phi)],\label{g_ev_0}
\ee
where $e$ is the electron charge, $\hbar$ Planck's constant, $\rhotip$ the density of states of the tip, assumed energy-independent, and $G_\psi$ is the local energy-dependent Green's function for the operator $\psi_0$. Using Eq.~\eqref{psicf} we have:
\be
G_\psi(\omega)=\tc^2 G_c(\omega) + \tf^2 G_f(\omega) + \tc \tf [G_{cf}(\omega) + G_{fc}(\omega)]\label{g_psi}
\ee
where the propagators and tunneling amplitudes have been assumed spin-independent, and the spin degeneracy is taken into account in the prefactor of Eq.~\eqref{g_ev_0}.
Abbreviating $\rho(\w) = -\I G(\w)/\pi$ the conductance is then given by
\be
g(e\Phi)=\frac{4\pi e^2}{\hbar} \rhotip[\tc^2 \rho_c +  \tc \tf (\rho_{cf} + \rho_{fc}) + \tf^2 \rho_f].\label{g_ev}
\ee
The mixed terms $\rho_{cf}$ and $\rho_{fc}$, arising from the interference of normal tunneling and cotunneling, typically induce a distinct p-h asymmetry at low energies, leading to a Fano-like lineshape for a single Kondo impurity and to an asymmetric double-peak structure surrounding the hybridization gap for a Kondo lattice.\cite{coleman09}


\section{Hybridization and interference terms}

Within the renormalized single-particle formulation,\cite{hewson} minimal bulk Hamiltonians for heavy-fermion metals as well as topological insulators involve four bands (i.e. two types of spin-degenerate orbitals) and can be written in Bloch form
\be
\mathcal{H}(\bk)=
\left(
\begin{array}{c|c}
 \epsilon_c(\bk) & V_{cf}(\bk)\\\hline
V_{fc}(\bk) & \epsilon_f(\bk)
\end{array}\label{h_general}
\right)
\ee
where all individual entries are $2\times 2$ matrices in spin (or pseudospin) space. Here, $\epsilon$ and $V$ denote kinetic-energy and hybridization terms, respectively, all of them representing interaction-renormalized quantities.\cite{hewson}

Assuming inversion symmetry, we have $\epsilon_c(\bk) = \epsilon_c(-\bk)$ and $\epsilon_f(\bk) = \epsilon_f(-\bk)$, while the hybridization can be even or odd, $V_{cf}(\bk) = \pm V_{cf}(-\bk)$.
In particular, TKIs with inversion symmetry \cite{tki1} are characterized by an odd hybridization function, typically arising from conduction-electron orbitals which are even under inversion ($s$ or $d$) and localized-electron orbitals which are odd ($f$).

In all cases, the orbital-off-diagonal propagators obey:
\bea
G_{cf}(\bk)&=&G^0_c(\bk)V_{cf}(\bk)G_f(\bk),\label{g_cf}\\
G_{fc}(\bk)&=&G_f(\bk)V_{fc}(\bk)G^0_c(\bk)\label{g_fc}
\eea
where $G^0_c(\bk)$ is the Green's function for $c$ electrons when $V=0$, and all propagators are taken at the same fixed frequency $\w$. Given that $G_c^0$ and $G_f$ are even under inversion, $G_{cf}$ and $G_{fc}$ inherit the inversion eigenvalues of $V_{cf}$ and $V_{fc}$, i.e., odd (even) hybridization implies odd (even) off-diagonal propagators.

The interference term in the tunneling conductance \eqref{g_ev} is determined by the \textit{local} off-diagonal propagator $G_{cf} \equiv G_{cf}(\br=\br')$ where translation invariance has been assumed. For an inversion-symmetric system with odd hybridization, this vanishes:
\bea
G_{cf}\equiv\int G_{cf}(\bk) d\bk=0,\label{g_cf_r}
\eea
as a consequence of the integrand being odd in $\bk$ -- this observation is central for the rest of the paper. Obviously $G_{cf}=0$ implies the absence of the interference term in Eq.~\eqref{g_ev}, and the tunneling lineshape needs to be re-considered. While this analysis strictly applies to a layered, i.e., effectively two-dimensional (2D), system where the surface layer probed by STS displays bulk properties, it must be modified for three-dimensional (3D) systems: Here, translation and inversion symmetries are broken by the surface, and an effective local $G_{cf}\ne 0$ is created as a consequence, such that the interference term of Eq.~\eqref{g_ev} reappears. However, its energy dependence is distinct from that in the case of even hybridization.


\section{Numerical examples}
To confirm and illustrate this reasoning, we have numerically studied a number of 2D and 3D Kondo-lattice models, the latter in a $(001)$ slab configuration to study the surface tunneling spectrum. In all cases, we work in a single-particle formulation with fixed renormalized parameters (i.e. ignoring self-consistency \cite{hewson}).
This implies that we neglect material-dependent surface effects, i.e., reduced Kondo temperature or Kondo breakdown \cite{kondo_breakdown} as well as surface reconstruction \cite{smb6_rec_io} and dangling bonds, polar charges, passivation, etc.

The examples include topologically trivial and non-trivial band structures, the latter corresponding to TKIs. Given the non-interacting character of the effective models, different electron concentrations can simply be achieved by tuning the overall chemical potential, and we will explicitly display figures for the Kondo-insulating cases.
For simplicity, we work on square (2D) or cubic (3D) lattices, with kinetic-energy matrices of the form
\begin{equation}
\epsilon_{c,f}(\bk)=(\epsilon_{c,f}-2t_{c,f} \sum_j\cos k_j) \mathds{1} \label{kin_en}
\end{equation}
where $j=x,y$ in 2D and $j=x,y,z$ in 3D. The nearest-neighbor (NN) hopping is taken as $t_c>0$, $t_f<0$; in the examples we will use $t_f/t_c=-0.1$.
Numerical values for $\epsilon_{c,f}$ and $V$ are chosen to ensure comparable bandwidths and hybridization gaps in all cases.


\begin{figure}[bt]
\includegraphics[width=0.49\textwidth]{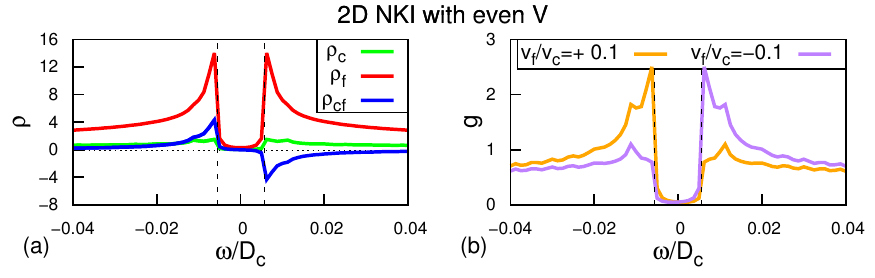}
\caption{
Numerical results for a 2D NKI with even hybridization \eqref{veven2d} at p-h symmetry, $\epsilon_c=\epsilon_f$.
(a) Imaginary parts of the Green's functions, $\rho_c$, $\rho_f$, and $\rho_{cf}=\rho_{fc}$.
(b) Conductance $g$ (in arbitrary units) for both signs of $\tf /\tc$.
The hybridization is $V=0.011 D_c$ where $D_c=8 t_c$ is the $c$-bandwidth.
The vertical dashed lines denote the bulk gap $\Delta/D_c=0.012$, and the small weight inside it is due to finite broadening $\delta/D_c=2.5\times 10^{-4}$ used in the calculation.
%
}\label{fig_even_2d}
\end{figure}

\subsection{2D model, even hybridization}
We start with simple 2D models representative for layered heavy-fermion materials. For the case of even hybridization we choose a local, spin-independent version,
\begin{equation}
V_{cf}(\bk)=V \mathds{1}
\label{veven2d}
\end{equation}
with $V<0$; we similarly take spin-independent tunneling amplitudes, $\tc\equiv \tc^\uparrow = \tc^\downarrow$, $\tf\equiv \tf^\uparrow = \tf^\downarrow$.
If the chemical potential is placed inside the hybridization gap, this yields a topologically trivial (or normal) Kondo insulator (NKI).
In this case $\rho_{cf}=\rho_{fc}$ is in general non-zero, peaked where most of the $f$ states live, i.e. just above and below the gap, and zero in the gap.
The sign of $\rho_{cf}$ is related to the fact that below (above) the gap the (anti)bonding linear combination of $c$ and $f$ states is created; in particular, at p-h symmetry, $\rho_{cf}$ is odd in $\omega$: $\rho_{cf}(-\omega)=-\rho_{cf}(\omega)$.
Concrete results are in Fig. \ref{fig_even_2d}, which illustrate an asymmetric tunneling signal arising from the interference term $\rho_{cf}$.

\subsection{2D model, odd hybridization}

\begin{figure}[!tb]
\includegraphics[width=0.49\textwidth]{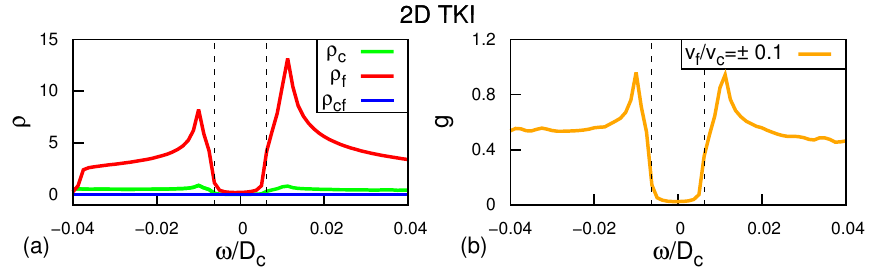}
\caption{
Same as Fig. \ref{fig_even_2d}, but for a 2D TKI with odd hybridization \eqref{v_cf_2d} close to p-h symmetry (exactly at p-h symmetry the system is metallic).
Since $\rho_{cf}$ is always zero, the lineshape is here independent of the sign of $\tf /\tc$. Parameters are $\epsilon_c-\epsilon_f=-0.125 D_c$, $V=0.026 D_c$.
}\label{fig_wti_2d}
\end{figure}

\begin{figure}[!bt]
\includegraphics[width=0.49\textwidth]{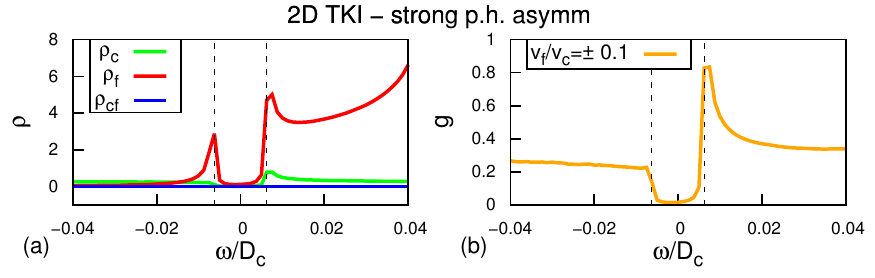}
\caption{Same as Fig. \ref{fig_wti_2d} for a 2D TKI, but with strong p-h asymmetry. This
causes an asymmetric lineshape without the need for an interference term ($\rho_{cf}=0$). Parameters are $V=0.04 D_c$, $\epsilon_c-\epsilon_f=-0.5 D_c$.
}\label{fig_wtib_2d}
\end{figure}
%
To demonstrate the effect of odd hybridization in a planar geometry, we choose a non-local, spin-dependent hybridization of the form
\begin{equation}
V_{cf}(\bk)=-2 i V(\sigma_x\sin k_x  + \sigma_y\sin k_y),\label{v_cf_2d}
\end{equation}
and $V_{fc}(\bk)\equiv [V_{cf}(\bk)]^\dagger=-V_{cf}(\bk)$ ($V_{cf}(\bk)$ is antihermitian).
This model realizes a 2D TKI.
Our specific choice of the relative phase between conduction and localized states (easily generalizable to 3D, see below) implies a factor of $i$ occurring in $V_{cf}$ \eqref{v_cf_2d}; moreover, the tunneling between $f$ orbitals and tip is along $z$ with amplitudes $\tf\equiv \tf^\uparrow = -\tf^\downarrow$ with $\tf$ being real, and $\tc\equiv \tc^\uparrow = \tc^\downarrow$.
The Green's functions remain diagonal in (pseudo-)spin space, $\rho^{\uparrow\downarrow}=\rho^{\downarrow\uparrow}=0$, and Eq.~\eqref{g_ev} for the tunneling conductance continues to apply, with
$\rho_{c}\equiv\rho_{c}^{\uparrow\uparrow}=\rho_{c}^{\downarrow\downarrow}$, $\rho_{f}\equiv\rho_{f}^{\uparrow\uparrow}=\rho_{f}^{\downarrow\downarrow}$, and $\rho_{cf,fc}\equiv\rho_{cf,fc}^{\uparrow\uparrow}=-\rho_{cf,fc}^{\downarrow\downarrow}$.

As emphasized above, $\rho_{cf}$ is strictly zero here, as demonstrated in Eq. \eqref{g_cf_r}, while
$\rho_c$ and $\rho_f$ are not qualitatively different from the even-hybridization case, compare Figs. \ref{fig_even_2d}(a) and \ref{fig_wti_2d}(a).
In the case of strong p-h asymmetry, Fig.~\ref{fig_wtib_2d}(a), the two peaks below and above the gap are strongly asymmetric.
For weakly coupled layers $\rho_{cf}$ will be non-zero but very small, such that the main conclusions remain unchanged.

\begin{figure}[tb]
\includegraphics[width=0.49\textwidth]{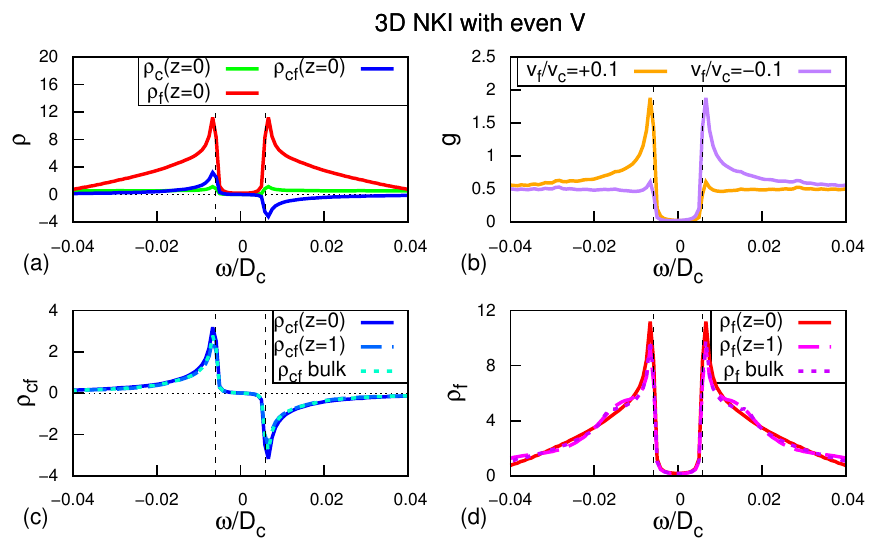}
\caption{
Numerical results for a 3D NKI with even hybridization \eqref{veven3d} at p-h symmetry, $\epsilon_c=\epsilon_f$, calculated in a slab geometry with $N\!=\!30$ layers.
(a) Imaginary part of the Green's functions, $\rho_c(z,\omega)$, $\rho_f(z,\omega)$, $\rho_{cf}(z,\omega)=\rho_{fc}(z,\omega)$ on the top layer $z\!=\!0$.
(b) Conductance $g$ for both signs of $\tf /\tc$.
(c) $\rho_{cf}(z,\omega)$ in different layers $z$, with $\rho_{cf}(z,\omega)=-\rho_{cf}(z,-\omega)$.
(d) $\rho_{f}(z,\omega)$ for different $z$.
Here $D_c=12 t_c$, $V=0.01 D_c$.
}\label{fig_even}
\end{figure}

\subsection{3D model, even hybridization}
For the 3D case, where STS can probe the surface but not the bulk, we start again with the case of even hybridization,
\begin{equation}
V_{cf}(\bk)=V \mathds{1}.
\label{veven3d}
\end{equation}
We assume that the tip only probes the topmost layer, i.e., in a slab calculation which yields $\rho(z)$ we take $\rho(z\!=\!0)$ when evaluating the tunneling signal according to Eq.~\eqref{g_ev}.
As in the 2D case we have $\tf\equiv \tf^\uparrow = \tf^\downarrow$ and $\tc\equiv \tc^\uparrow = \tc^\downarrow$.

The results from the slab calculation with even hybridization, Fig. \ref{fig_even}, show that the relevant quantities $\rho_c$, $\rho_f$, and $\rho_{cf}$ display a very weak dependence on the layer index $z$; moreover they are similar to the 2D case, Fig. \ref{fig_even_2d}.
As a consequence, neglection of surface effects is a reasonable approximation in this case, and a bulk theory is still appropriate to approximately describe the surface.
We recall that self-consistency has been neglected: this would take into account that Kondo screening is modified near the surface compared to the bulk, but would not change our conclusions concerning p-h asymmetries.

\subsection{3D model, odd hybridization}
We now turn to the case of odd hybridization in 3D. The simplest model \cite{tki1,tki2} for a TKI in 3D is obtained using a hybridization matrix
\bea
V_{cf}(\bk)=-2 i V(\sigma_x\sin k_x  + \sigma_y\sin k_y  + \sigma_z\sin k_z ),\label{v_cf}
\eea
and $V_{fc}(\bk)\equiv [V_{cf}(\bk)]^\dagger=-V_{cf}(\bk)$; we will take $V<0$.
The choice of relative phases of Wannier states yields real hybridization terms in real space along the $z$ direction.
As in the 2D case, $\tf\equiv \tf^\uparrow = -\tf^\downarrow$ and $\tc\equiv \tc^\uparrow = \tc^\downarrow$;
Upon modelling STS, we restrict our attention to a $(001)$ surface.

The model described by Eqs. \eqref{h_general}, \eqref{kin_en}, \eqref{v_cf} admits different phases depending on the energy difference $\epsilon_c-\epsilon_f$\cite{tki1,tki2}:
when $|\epsilon_c-\epsilon_f|>6(|t_c|+t_f)$, it realizes a topologically trivial insulator,
when $|\epsilon_c-\epsilon_f|<2(|t_c|+t_f)$, it realizes a weak topological insulator,
otherwise a strong topological insulator.

\begin{figure}[tb]
\includegraphics[width=0.49\textwidth]{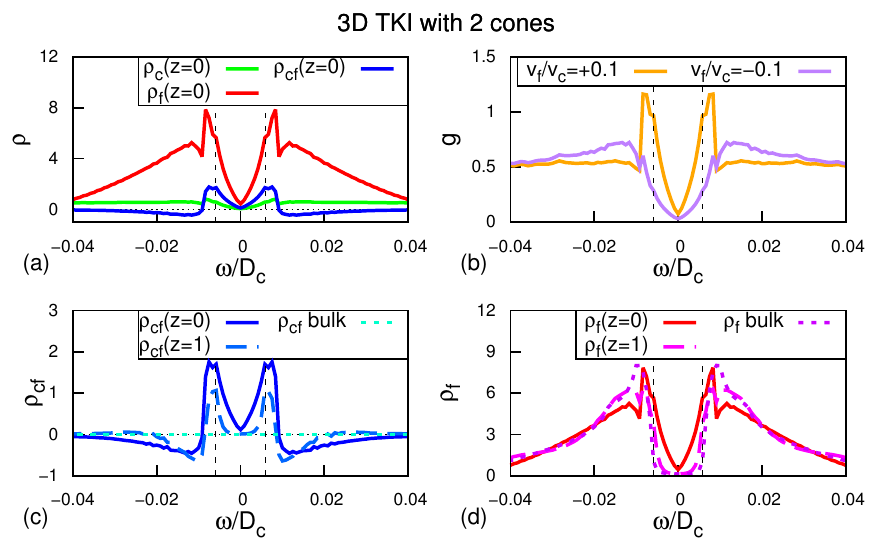}
\caption{
Same as Fig. \ref{fig_even}, but for a weak TKI with odd hybridization \eqref{v_cf} and two identical surface Dirac cones at p-h symmetry $\epsilon_c=\epsilon_f$; $V=0.021 D_c$.
We note that here $\rho_{cf}(z,\omega)=\rho_{cf}(z,-\omega)$;
and $\rho_{cf}$ depends strongly on the layer index, being zero in the bulk, showing that a bulk description of the surface is inadequate.
}\label{fig_wti}
\end{figure}

\begin{figure}[tb]
\includegraphics[width=0.49\textwidth]{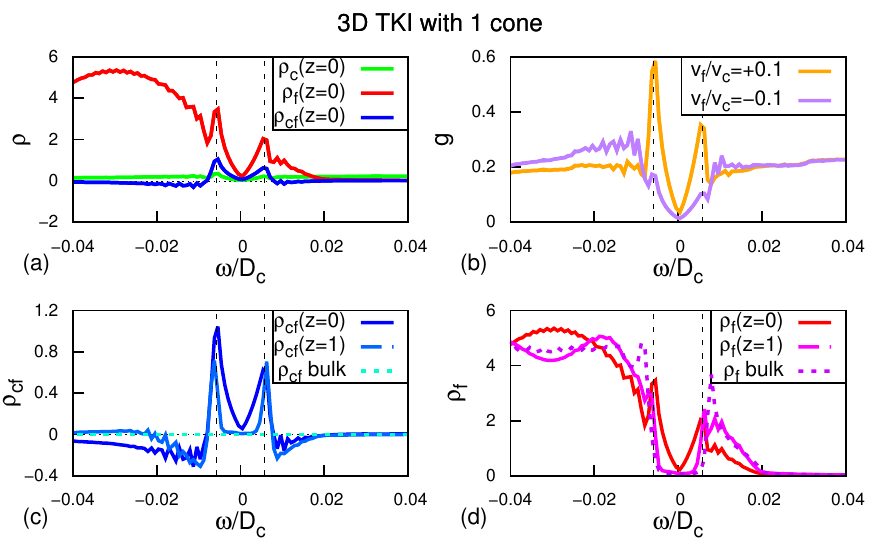}
\caption{
Same as Fig. \ref{fig_even}, but for a strong TKI with a single cone whose Dirac energy is in the bulk gap. Parameters are $V=0.021 D_c$, $\epsilon_c-\epsilon_f=0.33 D_c$; the small wiggles in the energy dependence arise from the finite thickness of the slab ($N=30$).
}\label{fig_odd}
\end{figure}

In Fig. \ref{fig_wti} we present results in the weak topological phase with two surface Dirac cones at the p-h-symmetric point ($\epsilon_c=\epsilon_f$), and in Fig. \ref{fig_odd}  in the strong topological phase with one surface Dirac cone away from p-h symmetry. In both cases, in-gap spectral weight appears due to the presence of topological surface states.
We find in general $\rho_{cf}(z,\omega)=\rho_{fc}(z,\omega)\ne 0$ unless $\rho_{cf}$ is measured deep in the bulk, Figs~\ref{fig_wti}(c) and \ref{fig_odd}(c). This may be interpreted as an effective local even-hybridization term which is {\em generated} by the presence of the surface.
Consequently, an interference term now appears in the tunneling conductance. However, its energy dependence is found to be completely different from the even-hybridization case in Fig. \ref{fig_even}.
Firstly, it has an appreciable weight in the gap arising from the topological surface states;
secondly, it changes sign not in the gap, but for energies corresponding to bulk $f$ states: this happens twice
for $\rho_{cf}(z=0,\omega)$, while $\rho_{cf}(z>0,\omega)$ has in general more nodes.

In particular, in the p-h-symmetric case of Fig. \ref{fig_wti}, $\rho_{cf}(z,\omega)$ is here an even function of energy,
$\rho_{cf}(z,-\omega)=\rho_{cf}(z,\omega)$,
while with even hybridization, as remarked, it is odd.
Even in absence of p-h symmetry, Fig. \ref{fig_odd}, the overall behavior of $\rho_{cf}(z,\omega)$ does not change.

\begin{figure}[tb]
\includegraphics[width=0.49\textwidth]{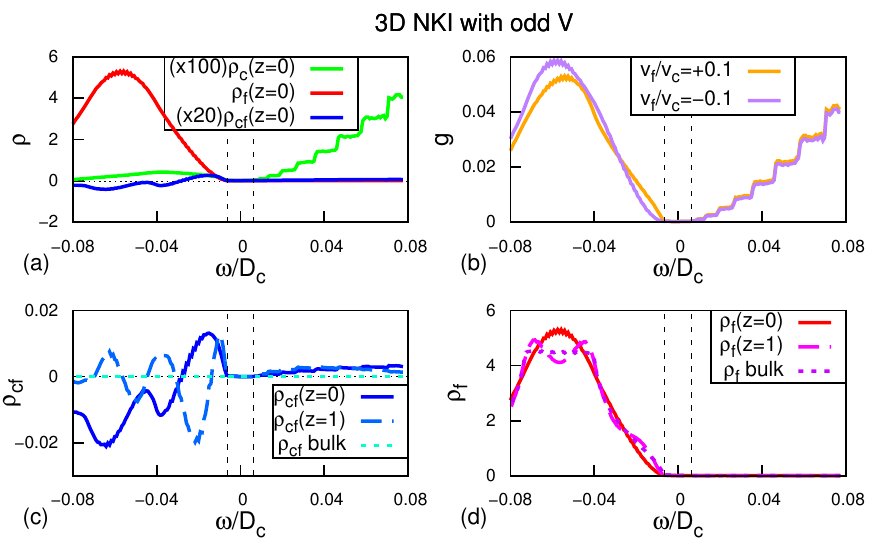}
\caption{
Same as Fig. \ref{fig_even}, but for the odd-hybridization model \eqref{v_cf} in the topologically trivial phase corresponding to a 3D NKI.
In this case $\rho_{cf}$ is non-zero on the surface, but extremely small when compared to Figs. \ref{fig_wti}, \ref{fig_odd}. Moreover, in-gap surface states are absent.
Parameters are $V=0.021 D_c$, $\epsilon_c-\epsilon_f=0.56 D_c$.
}\label{fig_bi}
\end{figure}

We have repeated the calculation in the topologically trivial phase of the model;\cite{tki1} 
note that in this case an insulator would be obtained even for zero hybridization, since the $c$ and $f$ bands do not intersect when $V=0$.
As shown in Fig. \ref{fig_bi},  we see that, given the same hybridization as in Fig. \ref{fig_odd}, $\rho_{cf}(z,\omega)$ is much smaller here, and it is zero in the gap due to the absence of surface states. 
However, $\rho_{cf}(z=0,\omega)$ changes sign in the $f$ band and not in the gap, in analogy to what happens in the topological non-trivial phase, and in contrast to the case with even hybridization.

Taking the information from Figs.~\ref{fig_wti}, \ref{fig_odd}, \ref{fig_bi} together, we can therefore attribute the occurrence of non-zero in-gap weight of $\rho_{cf}(z,\omega)$ to topological surface states, and speculate that its change of sign in the gap or in the $f$ band is due to even vs. odd hybridization.



\subsection{Toy model for {\sm}}

\begin{figure}[tb]
\includegraphics[width=0.49\textwidth]{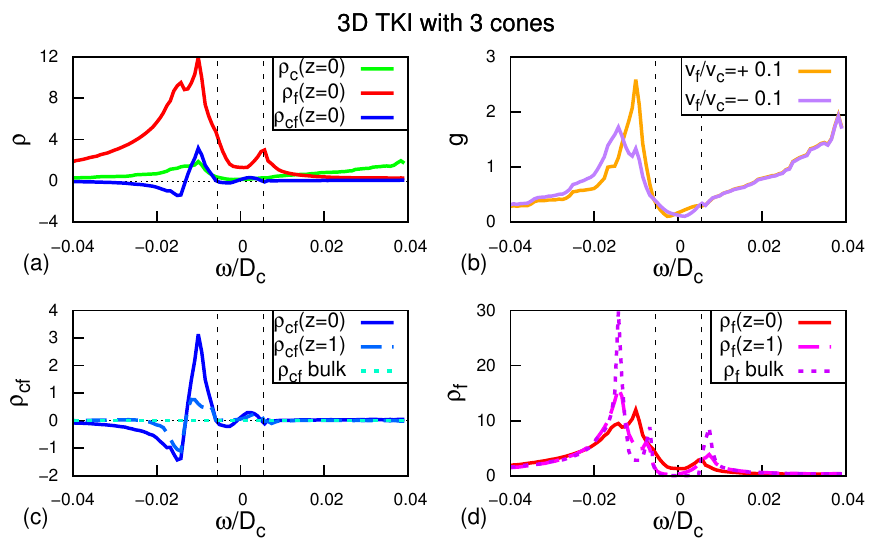}
\caption{
Same as Fig. \ref{fig_even}, but for a strong TKI with three Dirac cones and parameters $V=0.031 D_c$, $\epsilon_c-\epsilon_f=0.167 D_c$, $t_f'/t_f=-0.4$, $t_c'/t_c=-0.5$.
}\label{fig_smb}
\end{figure}

As a last example we provide results for a system with three Dirac cones, which should mimic {\sm}.\cite{sigrist_tki_prb}
To do this, we modify the kinetic energy by introducing 2nd NN hopping:
\begin{equation}
\epsilon_{c,f}'(\bk)=-4t_{c,f}' \sum_{j\ne l} \cos k_j \cos k_l \mathds{1} \label{kin_en_2},
\end{equation}
in a such a way to have a band inversion at the three $X$ points.\cite{sigrist_tki_prb}
Parenthetically, we note that this simplified model has a surface spin structure in disagreement with experiment \cite{smb6_arpes_mesot_spin} due to incorrect mirror Chern numbers,\cite{smb6_prl_io,sigrist_tki}
but this is largely irrelevant for the spin- and momentum-integrated STS spectrum considered here.
For this model, results are shown in Fig. \ref{fig_smb} where we can observe the same general features that we described for a TKI with one or two cones.
However, here one of the cones is inequivalent to the other two, with a different Dirac energy,
hence we do not clearly observe $\rho_{c,f,cf}(z,\omega)$ going to zero at the center of the cones.

\section{STS lineshape}

The above discussion has consequences for the energy dependence of the STS spectrum, i.e., the lineshape, and the role of the tunneling paths in Kondo systems.
In standard models for heavy-fermion materials, the phenomenological parameters $\tc$ and $\tf$ affect this lineshape:
their relative magnitude weighs the contributions from $\rho_c$, $\rho_f$ and $\rho_{cf}$,
and their relative sign tells if $\rho_{cf}$ enters with a plus or a minus sign, see Eq. \eqref{g_ev}, which in turn increases the strength of the peak either at positive or negative energies;
see Figs. \ref{fig_even_2d}(b), \ref{fig_even}(b).
In a TKI, instead, the relative sign enhances or depresses all features related to surface states; see Figs. \ref{fig_wti}(b), \ref{fig_odd}(b).

We note that, if the STM apex atom behaves like a $c$ orbital in the material,
we can approximate $\tf \approx V<0$, $\tc \approx -t_c<0$,
leading to $\sgn(\tf \tc)=\sgn (-t_c V)>0$.
In the STS signal this situation enhances both features below the gap in a NKI, Figs. \ref{fig_even_2d}(b), \ref{fig_even}(b),
and features from surface states in a TKI, Figs. \ref{fig_wti}(b), \ref{fig_odd}(b).

When $\rho_{cf}$ is small, for example in the case of the NKI with odd hybridization, Fig. \ref{fig_bi}(b), or for a TKI in 2D, Figs. \ref{fig_wti_2d}(b), \ref{fig_wtib_2d}(b)
the STS signal is mostly independent of the sign of $\tf/\tc$, and simply proportional to the sum
$\tc^2 \rho_c + \tf^2\rho_f$, i.e. the interference term can be neglected.
We stress that $\rho_{cf}=0$ does not always imply a symmetric lineshape, i.e., strongly p-h-asymmetric bands can lead alone to an asymmetric lineshape, Fig. \ref{fig_wtib_2d}(b). Conversely, in the 2D case with odd hybridization, p-h asymmetry is the only source of an asymmetric lineshape (assuming point tunneling).


\begin{figure}[!tb]
\includegraphics[width=0.49\textwidth]{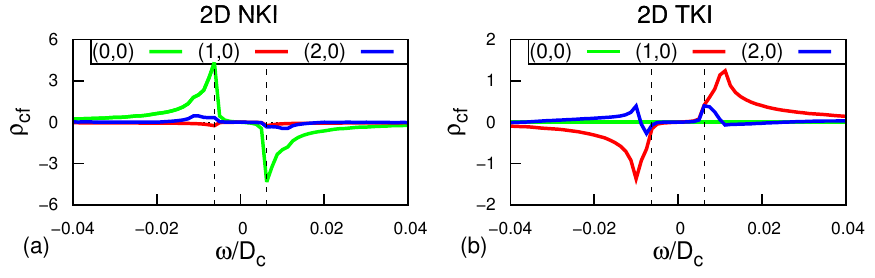}
\caption{
For the same parameters as in Figs. \ref{fig_even_2d} and \ref{fig_wti_2d}, we show the non-local $\rho_{cf}(\br-\br')$ as a function of the position $(\br-\br')\equiv(x,y)$:
(a) for even hybridization, $\rho_{cf}\equiv\rho_{cf}^{\uparrow\uparrow}=\rho_{cf}^{\downarrow\downarrow}$ is even in the position $\rho_{cf}(x,y)=\rho_{cf}(-x,-y)$, has a maximum on-site, at $(0,0)$, and decays for larger distances,
(b) for odd hybridization, $\rho_{cf}\equiv\rho_{cf}^{\uparrow\downarrow}=\rho_{cf}^{\downarrow\uparrow}$ is odd in the position $\rho_{cf}(x,y)=-\rho_{cf}(-x,-y)$, has a node on-site, at $(0,0)$, a maximum at $(\pm 1,0)$, $(0,\pm 1)$,
then decreases at larger distances
($\rho_{cf}^{\uparrow\uparrow}=\rho_{cf}^{\downarrow\downarrow}$ is here always zero).
%
}\label{fig_rrp}
\end{figure}


\section{Extension to more realistic models}

In this paper we have employed simple four-band models while real materials, like {\sm}, are described by more complicated models.\cite{tki_cubic,prbr_io_smb6}
We expect, however, that the main results of this paper remain valid as long as cubic (or tetragonal) symmetry is preserved.

Concerning {\sm} we remark that existing STS measurements \cite{pnas_smb6_stm,hoffman_smb6} find quite generally a Fano lineshape close to the Fermi energy.
This appears well described\cite{pnas_smb6_stm,hoffman_smb6} by existing simplified bulk theories \cite{coleman09,fano_lattice} -- where the asymmetry mainly arises from bandstructure effects --
except for the residual conductance in the (pseudo)gap, which is due to some other conduction channel.
A topological bandstructure model, such as ours, naturally assigns this residual conductance to topological surface states (but details are beyond our simplified description).

The tunnelling theory used in this paper assumed point-like tunneling between tip and surface. If the tip is larger, we have to consider the non-local Green's function $G(\br,\br')$, which can be non-zero even when $G(\br,\br)=0$;
hence, our approach is only a starting approximation of the phenomenon.
As a general rule, in presence of translational symmetry $G_{cf}(\br-\br')$ will have a node when $(\br-\br')=0$,
in contrast to the even-hybridization case.
In Fig. \ref{fig_rrp} we show how $\rho_{cf}$ depends on $(\br-\br')$ in 2D, with either even or odd hybridization.

Finally we note that a fully accurate description of the tunneling signal requires the explicit use of the spatial shape of the Wannier functions \cite{kreisel_stm},
which, however, implies more complicated numerical work which is beyond the scope of this paper.


\section{Conclusions}
For heavy-fermion systems with bulk inversion symmetry we have argued that a hybridization term which is odd under inversion leads to the vanishing the local non-diagonal term $G_{cf}$ of the Green's function. We have shown that this strongly modifies considerations concerning the spectral lineshape as measured in STS experiments, because the cotunneling interference term usually made responsible for the appearance of a Fano-like lineshape is either absent or very different from that in the even-hybridization case.
Sharp conclusions can be drawn for p-h-symmetric bands, where even hybridization generically leads to a p-h-asymmetric STS signal, its asymmetry being entirely due to $G_{cf}$, and odd hybridization to a strictly p-h-symmetric lineshape.
In a general p-h-asymmetric situation, an asymmetric STS lineshape can arise from both band asymmetries and tunneling interference, and additional information (e.g. from ARPES or bandstructure calculations) is required to deduce properties of tunneling paths or hybridizations from STS data.
We have provided concrete examples for simple models of topological and non-topological Kondo insulators which we hope to be useful for the interpretation of future tunneling experiments.

\acknowledgments

We thank J. E. Hoffman, L. Fritz, A. Mitchell, D. K. Morr, and S. Wirth for discussions and collaborations on related work.
This research was supported by the DFG through SFB 1143 and GRK 1621
as well as by the Helmholtz association through VI-521.

\bibliographystyle{apsrev4-1}
\bibliography{tki}

\end{document}